\documentstyle[prl,aps,epsf]{revtex}
\draft

\begin{document}
\twocolumn[\hsize\textwidth\columnwidth\hsize\csname @twocolumnfalse\endcsname

\title{Density of States Approach to Electric Field Fluctuations in Composite Media  }
\author{D. Cule and S. Torquato}
\address{Princeton Materials Institute, Princeton University, 
         Princeton, New Jersey 08544}
\date{\today }
\maketitle

\begin{abstract}
Spatial fluctuations of the local electric field induced by a constant 
applied electric field in composite media are studied analytically and
numerically. It is found that the density of states for the fields
exhibit sharp peaks and abrupt changes in the slope at certain critical 
points which are analogous to van Hove singularities in the density of
states for phonons and electrons in solids. As in solids, these 
singularities are generally related to saddle and inflection points in the field
spectra and are of considerable value in the characterization
of the field fluctuations. The critical points are very prominent in dispersions with a
regular, ``crystal-like'',  structure. However, they broaden and
eventually disappear as the disorder increases.

\vspace{10pt}

PACS number(s): 05.40.+j, 61.43.-j, 77.90.+k
\vspace{10pt}
\end{abstract}
] 

In the study of heterogeneous materials, the preponderance of work has been 
devoted to finding the effective transport, electromechanical and mechanical properties of the 
material~\cite{Ch79}, which amounts to knowing only the first moment of the 
local field. When composites are subjected to constant applied fields, 
the associated local fields exhibit strong spatial fluctuations. The analysis 
and evaluation of the distribution of the local field has received far less 
attention. Nonetheless, the distribution of the local field is of great 
fundamental and practical importance in understanding many crucial material 
properties such as breakdown phenomenon~\cite{Li89} and the nonlinear behavior 
of composites~\cite{Le94}. Much of the work on field distributions has been 
carried out for lattice models using numerical~\cite{De86,R5} and  perturbation 
methods \cite{R6}. Recently, continuum models have been also addressed using 
numerical techniques~\cite{R7}.

In this Letter, we study the local electric field fluctuations by
analyzing the density of states for the fields.
To illustrate the procedure, we evaluate the
density of states for three different continuum models of dielectric 
composites: the Hashin-Shtrikman (HS) construction~\cite{R1}, periodic and
random arrays of cylinders. 
It is found that the density of states for the fields  exhibits sharp peaks and abrupt changes in the slope 
at certain critical points which are analogous to van Hove singularities in 
the density of states for phonons and electrons in solids. 
This analogy is new and powerful, and places the study
of field fluctuations in composites on the firm foundation
of solid-state theory. In the case of the Hashin-Shtirkman
construction, we obtain an exact analytical expression
for the density of states. We first describe the basic
equations and then determine the density of states for
the aforementioned examples.

Consider a composite material composed of $(n-1)$ isotropic
inclusions with dielectric constants $\epsilon_i$ and volume fractions
$\phi_i$ ($i=2,\,\ldots ,n$) in a uniform reference matrix of dielectric constant 
$\epsilon_1$ with volume fraction $\phi_1$.  
Clearly, the local dielectric constant at position ${\bf r}$ is 
$\epsilon({\bf r}) = \sum_{i=1}^{n}\epsilon_i I^{(i)}({\bf r})$, where
$I^{(i)}({\bf r})$ is the characteristic function of phase $i$ which
has non-vanishing value $I^{(i)}({\bf r})=1$ only if ${\bf r}$  lies inside
the volume $V_i$ occupied by phase $i$.
Let ${\bf E}_0({\bf r})$ denote an applied electric field. The local electric
field ${\bf E}({\bf r})$, and the dielectric displacement ${\bf D}({\bf r})$ are related via
the relation
${\bf D}({\bf r})= \epsilon({\bf r}) {\bf E}({\bf r})$.
The potential field $u({\bf r})$ is related to $\bf E$ by
${\bf E}({\bf r}) = -\nabla u({\bf r})$. The local fields
are obtained from the solution of the governing relation
$\nabla\cdot {\bf D}({\bf r})= 0$ subject to appropriate
boundary conditions.

Since the local dielectric constant of the composite material is a 
piecewise continuous function, we can solve the following
equivalent equations:
\begin{eqnarray}
& &\nabla^2u({\bf r})= 0,\qquad\qquad\qquad\quad\;\; {\bf r} \in V_i,
\label{E2}
\\
& & u({\bf r})|_{{\bf r}^-} =  u({\bf r})|_{{\bf r}^+} ,
    \qquad\qquad\quad  {\bf r} \in \partial V_i, 
\label{E3}
\\
& &\epsilon_i \partial_n  u({\bf r})|_{{\bf r}^-}
   = \epsilon_j \partial_n  u({\bf r})|_{{\bf r}^+},  
\quad\; {\bf r} \in \partial V_i,
\label{E4}
\end{eqnarray}
where  $i=2,\,\ldots, n$. The index $j$ denotes 
neighboring phases in contact with a given inclusion $i$. We donote the outward normal 
derivative to the interface $\partial V_i$ by $\partial_n$ and 
the interface points approached from inside or outside inclusion by ${\bf r}^-$, 
and ${\bf r}^+$, respectively. The numerous interfaces between the inclusions 
and matrix are, in general, irregular and randomly distributed in space. 
In most cases solutions of (\ref{E2})-(\ref{E4}) can be found only numerically.

In order to show the salient features of the field distribution,
we first consider an analytically tractable model of composite media:
the HS composite-cylinder construction~\cite{R1}. Although the effective dielectric constant
$\epsilon_e$ of this model is known exactly, its local field
distribution has  heretofore not been investigated.
The HS two-phase model is made up of composite
cylinders consisting of a core of dielectric constant $\epsilon_2$ and
radius $a$, surrounded by a concentric shell of dielectric constant
$\epsilon_1$ and radius $b$. The ratio $(b/a)^2$ equals the
phase 2 volume fraction $\phi_2$ and the composite cylinders fill all
space, implying that there is a distribution in their sizes ranging 
to the infinitesimally small. 
For this special construction, it is enough to consider the electric
field within a single
composite cylinder in a matrix having the effective dielectric constant
$\epsilon_e = \epsilon_1[1-2a^2\beta/(a^2\beta +b^2)]$, where
$\beta = (\epsilon_2-\epsilon_1)/(\epsilon_2 + \epsilon_1)$.
Let the constant applied field
point in the $x$-direction, ${\bf E}_0({\bf r}) = E_o{\bf{\hat x}}$. 
Under this condition, the presence of the composite cylinder does not
change the distribution of the fields in the composite for $r > b$ nor the total
energy stored in the region occupied by the cylinder.

Within a cylindrical inclusion, the solution of (\ref{E2})-(\ref{E4}) with 
the boundary condition $u({\bf r}) = -E_0 r\cos (\theta)$ reads
\begin{eqnarray}
u({\bf r}) &=&\left\{
\begin{array}{l}
A r\cos(\theta), \qquad\qquad \mbox{for}\quad  r\le a,
\\
(B r + \frac{C}{r})\cos(\theta),\quad \mbox{for}\quad  a\le r\le b.
\end{array}
\right.
\label{E5}
\end{eqnarray}
\noindent
Consequently, the magnitude of the electric field is constant 
$|{\bf E}({\bf r})| = |A|$ for $r\le a$, and 
\begin{equation}
|{\bf E}({\bf r})| = |B|\sqrt{ 1 + (\frac{a}{r})^4\beta^2
+2\beta(\frac{a}{r})^2\cos(2\theta)}
\label{E6}
\end{equation}
\noindent
for $a\le r\le b$. The coefficients $A$,$B$, and $C$ depend on the 
geometry and material properties:
\begin{eqnarray}
 A &=& E_0(1-\beta)/(\beta\phi_2-1),
\label{E7} \\
 B &=& E_0/(\beta\phi_2-1),
\label{E8}\\
 C &=& E_0 a^2 \beta/(\beta\phi_2-1).
\label{E9}
\end{eqnarray}
\noindent

In the study of electric-field fluctuations, it is convenient to introduce a
density of states per unit volume, $g(E)$, defined so that $g(E)dE$ is the
total number of states in the range between $E$ and $E+dE$, divided
by the total volume of the inclusion:
\begin{equation}
g(E) = \frac{1}{\pi b^2}\int \delta(E-|{\bf E}({\bf r})|)d{\bf r}.
\label{E10}
\end{equation}
\noindent
Substituting (\ref{E6}) into (\ref{E10}) yields
\begin{eqnarray}
g(E) &=& \phi_2\delta(E-|A|) \nonumber
\\
&+& \frac{2\phi_2 E}{\pi B^2}\int_{\phi_2}^{1}dx
\frac{1}{x^2\sqrt{\gamma_E(x)}}\Theta\big(\gamma_E(x)\big),
\label{E11}
\end{eqnarray}
\noindent
where $\Theta(x)$ is the Heaviside step function, and
\begin{equation}
\gamma_E(x) =4(\beta x)^2-[1+(\beta x)^2 -(E/B)^2]^2.
\label{E12}
\end{equation}
\noindent
We note that $\gamma_E(x)$ is bounded from above, and that the integrand in Eq. 
(\ref{E11}) is non-zero only if $\gamma_E(x)\ge 0$,  implying that the 
density of states $g(E)$ is non-zero only for fields
in the finite {\em bandwidth}: $E\in[E_{\mbox{min}},E_{\mbox{max}}]$. 
The extremal field values for the problem at hand are
$E_{\mbox{min}}=|B|(1-|\beta|)$, and $E_{\mbox{max}}=|B|(1+|\beta|)$.

\begin{figure}
\narrowtext
\epsfxsize=3.0truein
\centerline{\epsffile{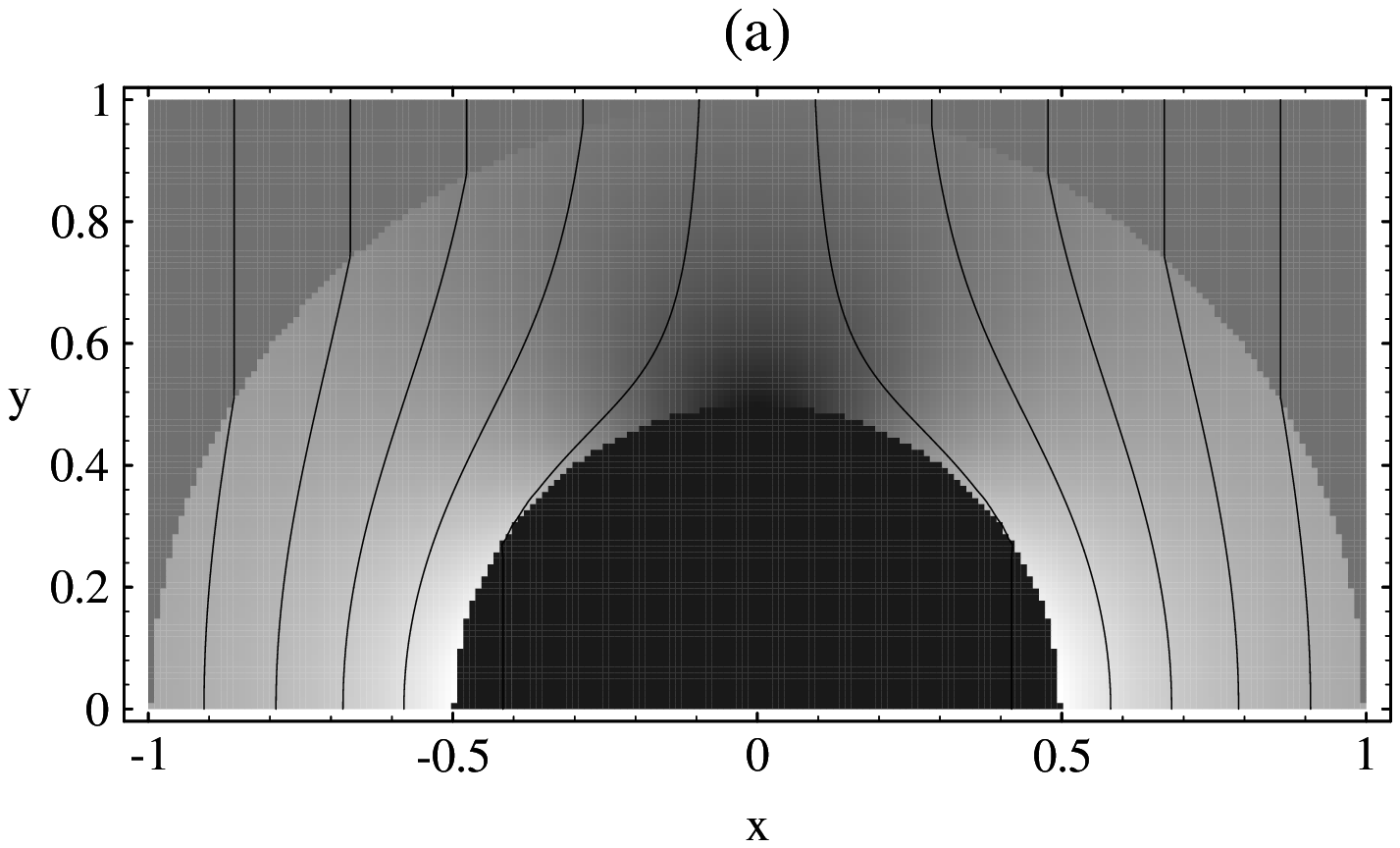}}
\vspace{\baselineskip}
\epsfxsize=3.0truein
\centerline{\epsffile{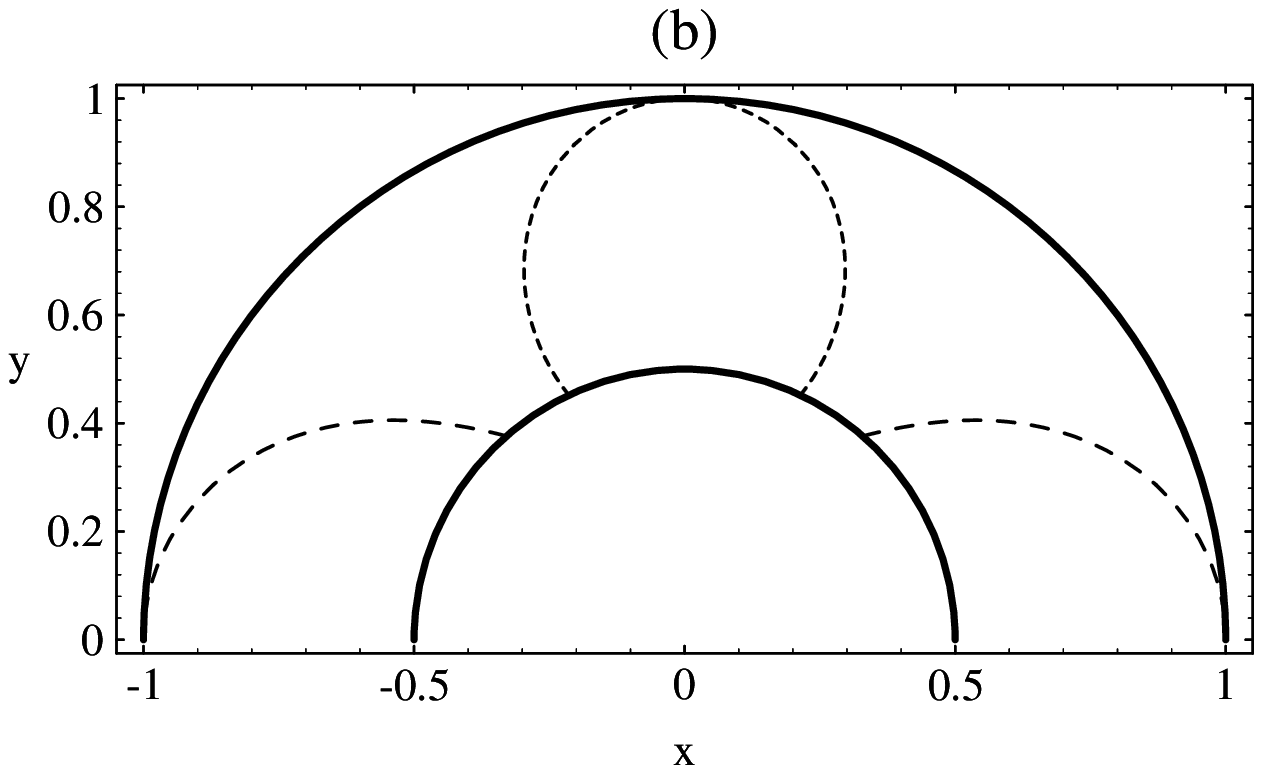}}
\vspace{\baselineskip}
\caption{
The upper half of a composite cylinder of outer radius $b=1$ with inner
core of radius $a=0.5$ (i.e., $\phi_2=(a/b)^2=0.25$), dielectric constant $\epsilon_2=10$, and
outer concentric shell of dielectric constant $\epsilon_1=1$:
(a) Electric field density plot in which lighter shades (gray-scale
representation) correspond to increasing field magnitudes, and
associated contours of equipotential lines $u({\bf r}) =$ const.
(b) Distribution of fields contributing to van Hove singularities
in the density of field states $g(E)$. Dotted lines show positions of
$E_{VH}=|B||1-\beta \phi_2|=E_0$ while dashed lines identify $E_{VH}=|B|(1+\beta \phi_2)$.
The thick lines denote the boundaries of the cylinders.
}
\label{F1}
\end{figure}

\begin{figure}
\narrowtext
\epsfxsize=3.0truein
\centerline{\epsffile{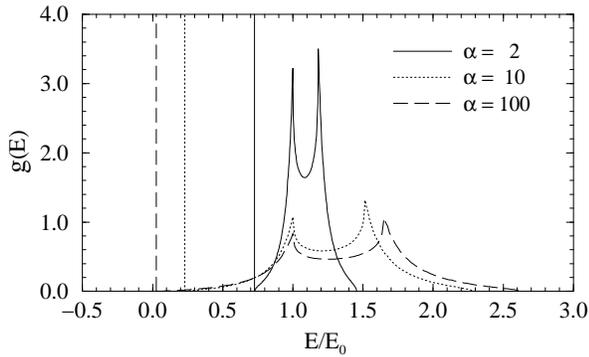}}
\vspace{\baselineskip}
\caption{
Density of  states $g(E)$ for the HS construction with
at $\phi_2=0.25$ for contrast values
$\alpha=\epsilon_2/\epsilon_1=$ 2, 10, and 100.
}
\label{F2}
\end{figure}

The highest field determines the electrical breakdown properties. 
Equally important
 are the fields at which $g(E)$ has its
maxima, i.e.,  the fields that occur  most frequently in the composite.  
To identify them, note that the dominant contributions to the integral in 
(\ref{E11}) come from regions close to the points 
$x_0  =|\beta|^{-1}|1\pm|E/B||$ which are solutions of  $\gamma_E(x_0)=0$.
Expanding $g(E)$ about $x_0$, and integrating (\ref{E11}) yields
$g(E)\sim (x-x_0)^{1/2}|_{x_0=\phi_2}^{1}$ or
$g(E)\sim[(1-x_0)^{1/2}-(\phi_2-x_0)^{1/2}]$. Thus, the density of states
$g(E)$ is not an analytic function since its derivatives are singular at 
$E=E_{VH}$ for which $x_0=1$ or $\phi_2$:
\begin{eqnarray}
E_{VH} &=&\left\{
\begin{array}{l}
E_{\mbox{min}},
\\
|B||1-\beta \phi_2|,
\\
|B||1+\beta \phi_2|,
\\
E_{\mbox{max}}.
\end{array}
\right.
\label{E15}
\end{eqnarray}
These singularities are very similar to the van Hove singularities 
found in the density of states for phonons and electrons in solids~\cite{R2}.
As in the case of solids, the singularities in  $g(E)$ are gernerally 
associated with  
saddle and inflection points on the surface of generated fields 
$|{\bf E}({\bf r})|$. At these points,  $g(E)$ exhibits  characteristic sharp 
local maxima or minima, with abrupt changes in the slope. 
This is illustrated in Figs.~\ref{F1} and ~\ref{F2}. Because 
the HS construction lacks translational symmetry (unlike the
subsequent period example), only  inflection points are present
\noindent
in the density
plot of $|{\bf E}({\bf r})|$ shown in Fig.~\ref{F1}(a). They occur in phase 1
along the lines $\theta=0$ and $\theta=\pi/2$.

Figure~\ref{F1}(a) shows the density plot of local electric fields together
with the contours of the equipotential lines. Inside $r\le a$ region, the
magnitude of electric field is constant. This generates the $\delta$-function
in the density of states (\ref{E11}).
To locate spatial coordinates of the singular fields $E_{VH}$, we first
consider the case $\epsilon_2 > \epsilon_1$. From Eq.~(\ref{E6}) it follows
that $E_{VH}=E_{\mbox{min}}$ where $(r=a,\theta=\pi/2)$.
Since $E_{\mbox{min}} = |A|$, the position of this van Hove singularity
overlaps with position of the $\delta$-function in $g(E)$.
The highest electric fields are expected to occur at the two-phase interface.
Inserting $E_{VH}=E_{\mbox{max}}$ into Eq.~(\ref{E6}) gives the corresponding 
coordinates $(r=a^+,\theta=0)$ and $(r=a^+,\theta=\pi)$. Notice that at $r=a$, 
the surface $|{\bf E}({\bf r})|$ has finite jump due to the discontinuity in 
normal component of $|{\bf E}|$ given by (\ref{E4}).

The locations of the singularities $E_{VH}=|B||1\pm\beta \phi_2|$ are not 
obvious.  In this case, the solution of (\ref{E6}) are the  curves
\begin{equation}
\theta =\pm \arccos\big\{ \frac{1}{2}\beta[\frac{a^2}{r^2}-\frac{a^2r^2}{b^4}]
-(\pm)\frac{r^2}{b^2}\big\} + k\pi,
\label{E16}
\end{equation}
\noindent
$k=0,\pm1,\pm2,\ldots\;$, and $r\in[a,b]$ which are plotted in
Fig.~\ref{F1}(b). The dotted central curves correspond to
$E_{VH}=|B||1-\beta \phi_2|=E_0$. The field strength on the
dashed curves is  $E_{VH}=|B||1 + \beta \phi_2|$. 

The case in which  $\epsilon_2 < \epsilon_1$ can be studied using
similar considerations.
For example, the peak of the $\delta$-function will be at 
$E_{VH}=E_{\mbox{max}}$ since the highest fields are always 
generated in the less conducting phase.

In Fig.~\ref{F2}, the density of states $g(E)$ of the composite cylinder 
model is plotted for several values of the contrast ratio 
$\alpha=\epsilon_2/\epsilon_1$, at fixed volume fraction $\phi_2= (a/b)^2=$
0.25. Increasing the contrast between inclusions leads to a
broadening of $g(E)$. With an increase of $\alpha$, the
maximum field strength (in the less conducting phase) increases while in the 
higher conducting phase, the amplitude of the constant field decreases.
The positions of van Hove singularities are easily identified as the
minimum or maximum field, or local maxima in $g(E)$. In the opposite 
limit, $\alpha\rightarrow 1$, the bandwidth of the allowed fields collapses 
to the single peak $\sim\delta(E-|A|)$, as expected.

\begin{figure}
\narrowtext
\epsfxsize=3.0truein
\centerline{\epsffile{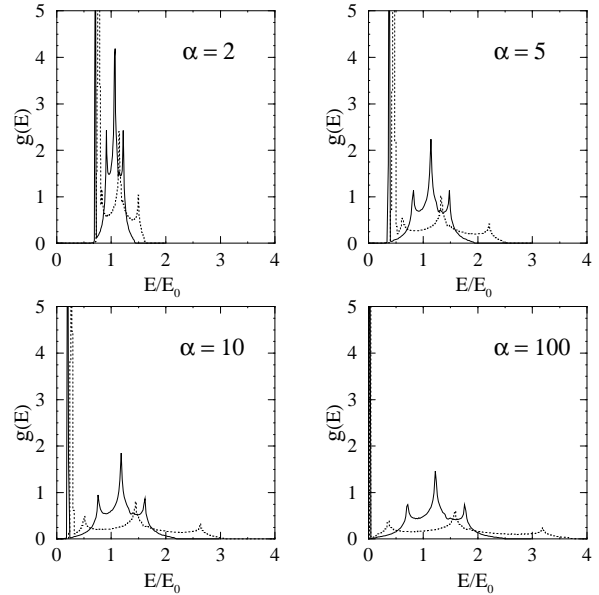}}
\vspace{\baselineskip}
\caption{
Density of field states $g(E)$ for square arrays of cylinders
for different contrasts $\alpha=\epsilon_2/\epsilon_1$, and different
volume fractions: $\phi_2\approx$ 0.2 (full lines),
0.4 (dotted lines). Potential and electric fields are calculated by
the finite difference method with resolution $L/400$, $L=1$.
}
\label{F3}
\end{figure}

Next we extend our considerations to composite material with periodic
microstructure. 
In particular, we consider a square array of cylinders of
dielectric constant $\epsilon_2$ in a matrix with dielectric constant $\epsilon_1$:
a model whose effective dielectric constant has been well studied
\cite{R3,R4}.
Let the distance between the centers of neighboring
cylinders be $2L$ and assume that the external field ${\bf E}_0$ is
applied along the x-axis.
Although analytical multipole expansion techniques
(leading to an infinite set of linear equations which must be truncated)
yield accurate estimates of the effective dielectric constant, they are not
adequate to obtain the field distribution.
Accordingly, we use the finite difference scheme to solve 
(\ref{E2})-(\ref{E4}).

The essence of the numerical method is to map the composite to a network of
conductors in which each conducting bond has the dielectric constant of the corresponding
region in the composite. The potential fields at each internal node are solutions 
of the system of equation $\sum_j\epsilon_{i,j}(u_i-u_j)=0$, where index $j$
runs over the nodes which are the nearest neighbors of the node $i$, where
$\epsilon_{i,j}$ is the dielectric constant of the bond between $i^{\mbox{th}}$ and
$j^{\mbox{th}}$ node. In addition to these equations, macroscopic boundary
conditions are imposed.
Because of periodicity, it is enough to consider the solution in a square box 
of size $L$ with a cylinder centered at one of its corners. Then the boundary 
condition on the edges along the direction
transverse to the applied field, say the $y$-direction, is
$\partial_y u(x,y)=\partial_y u(x,y+L)=0$. In the direction of the applied field,
there is finite gradient $u(x+L,y)-u(x,y)=-E_0$.

The solutions for the density of states $g(E)$ are shown in Fig.~\ref{F3}.
First we notice that in addition to the $\delta$-function-like peak associated 
with the fields inside the cylinders, there are three prominent
local maxima. These maxima are  signatures  of van Hove singularities. (Further details
about the topological properties of the field surfaces will be given elsewhere~\cite{Di98}.)
 The field fluctuations lie in the bandwidth of the density of states. 
Figure~\ref{F3} illustrates how by increasing
the phase contrast $\alpha$, the fluctuations grow. It seen that the same effect
is found if the volume fraction $\phi_2$ increases toward
the percolation threshold.

As a final example, we consider random arrays of cylinders whose density
of states is distinctly different than the first two examples. Previous
work on field fluctuations  reveal distributions with  
two {\em global} (as opposed to local) peaks \cite{R5,R7},
although they arose for different reasons in these two studies.
The presence of two global peaks is readily understood from our study as is
explained below.
In fact, we conjecture that random multiphase 
composites with widely different dielectric constants  will have 
many well separated global peaks. 

\begin{figure}
\narrowtext
\epsfxsize=3.0truein
\centerline{\epsffile{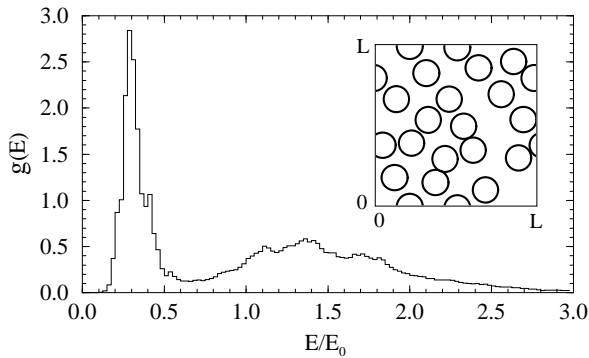}}
\vspace{\baselineskip}
\caption{
Density of states for a single realization of a random dispersion of nonoverlapping
cylinders (see inset) with $\alpha=10$, $\phi_2\approx0.4$, calculated by a finite
difference scheme with a resolution $L/400$.}
\label{F4}
\end{figure}

 Figure~\ref{F4} shows the histogram of $g(E)$ that we have calculated for a single 
realization of a random distribution of 20 nonoverlapping disks of dielectric 
constant $\epsilon_2=10$ in a matrix of $\epsilon_1=1$. The system size is $L=1$,
so that the volume fraction $\phi_2\approx 0.4$. Neglecting the irregularities
due to insufficient statistics in the number of calculated fields, the positions 
of the three local maxima (van Hove singularities) at 
$|{\bf E}({\bf r})|/|{\bf E}_0|\ge 1$ are still evident, although significantly 
diminished relative to the previous periodic example. 
Averaging~\cite{note} the density of states over many realizations of random samples, 
as was done in Refs. \cite{R5,R7}, additionally  smears the sharpness 
of the band edges and maxima. Therefore,  
disorder causes the local maxima to broaden and eventually  merge together, 
i.e., the local features disappear as the disorder increases. 
Similar disorder effects are well known in the theory of elecronic density of 
states in solids.

To summarize, we find van Hove type singularities in the density of states
for  local electric fields induced by an applied  field in composites. We show 
how these singularities are related to the maxima and the bandwidth of $g(E)$.
The complex multiple-peak behavior in the density of states are not adequately 
characterized by straightforward calculation of their lower moments. 
Analysis of $g(E)$ at its van Hove singular points represents a powerful new
approach to quantify field fluctuations in composite media.

It is noteworthy that the density of states analysis of field
fluctuations laid out in this paper for dielectric composites
can be applied to other field phenomena, including strain
fields in elastic media, velocity fields for flow through porous
media, and concentration fields for diffusion and reaction
in porous media.

We are grateful to  E.J. Garboczi for sharing his code on the
finite difference method with us. This work was supported by the
U.S. Department of Energy, OBES, under Grant No. DE-FG02-92ER14275.


\begin{references}
\bibitem{Ch79}
R. M. Christensen,
{\em Mechanics of Composite Materials},
(Wiley, New York, 1979); G. W. Milton and N. Phan-Thien,
Proc. R. Soc. Lond. A {\bf 380}, 305 (1982);
S. Torquato, Appl. Mech. Rev., {\bf 44}, 37 (1991).


\bibitem{Li89} 
Y. S. Li and P. M. Duxbury, Phys. Rev. B {\bf 40}, 4889 (1989).



\bibitem{Le94} 
O. Levy and D. J. Bergman, Phys. Rev. B {\bf 50}, 3652 (1994).

\bibitem{De86} 
L. de Arcangelis, S. Redner and A. Coniglio,
Phys. Rev. B {\bf 34}, 4656 (1986).

\bibitem{R5} Z. Chen and P. Sheng, Phys. Rev. B {\bf 43}, 5735 (1991);
Phys. Rev. Lett. {\bf 60}, 227 (1988).

\bibitem{R6} M. Barth\'{e}l\'{e}my and H. Orland, Phys. Rev. E {\bf 56}, 
2835 (1997).

\bibitem{R7} H. Cheng and S. Torquato, Phys. Rev. B {\bf 56}, 8060 (1997).

\bibitem{R1} Z. Hashin and S. Shtrikman, J. Appl. Phys. {\bf 33},
3125 (1962).

\bibitem{R2} L. van Hove, Phys. Rev. {\bf 89}, 1189 (1953).

\bibitem{R3} Lord J.W.S. Rayleigh, Phil. Mag. {\bf 34}, 481 (1892).

\bibitem{R4} W.T. Perrins, D.R. McKenzie and R.C. McPhedran,
Proc. R. Soc. Lond. A {\bf 369}, 207 (1979).


\bibitem{note}
Sample averaging is equivalent to calculaton  of  $g(E)$ for
a sample with a large number of inclusions. 

\bibitem{Di98}
D. Cule and S. Torquato, to be published.
\end{references}
\end{document}